\documentclass[prl,twocolumn,url]{revtex4}
\usepackage[page]{appendix}
\usepackage{graphicx,epsf,epsfig,amsmath}
\usepackage{alltt,dsfont,amsmath,amssymb,bm}
\usepackage{float}

\usepackage{hyperref}
\usepackage{color}

\newcommand{\iden}{ \mathds{ 1}}

\newcommand{\G}{{\cal{G}}}

\newcommand{\GH}{{\bf g}}
\newcommand{\GHI}{\GH^{-1}}
\renewcommand{\L}{{\cal L}}
\newcommand{\LL}{{\bf L}}

\newcommand{\si}{\sigma}
\newcommand{\sib}{\bar{\sigma}}

\newcommand{\tJ}{\ $t$-$J$ \ }

\newcommand{\U}{{\ \cal U}}

\newcommand{\V}{{\cal V}}

\newcommand{\bb}[1]{{{\mathbf #1}}}

\newcommand{\nn}{\nonumber}
\newcommand{\chem}{{\bm \mu}}

\newcommand{\beq}{\begin{equation}}
\newcommand{\eeq}{\end{equation}}
\newcommand{\barray}{\begin{eqnarray}}
\newcommand{\earray}{\end{eqnarray}}
\newcommand{\disp}[1]{Eq.~(\ref{#1})}
\newcommand{\refdisp}[1]{Ref.~(\onlinecite{#1})}
\newcommand{\figdisp}[1]{Fig.~(\ref{#1})}
\newcommand{\factor}{\frac{1}{\sqrt{\Omega}}}
\newcommand{\D}{{\cal D}}

\newcommand{\sw}{a_{\cal G }}

\newcommand{\lab}[1]{\label{#1}}

\newcommand{\llangle}{\langle \langle}
\newcommand{\rrangle}{\rangle \rangle}

\begin{document}
\title{ Extremely Correlated Fermi Liquid study of 
the $U= \infty$   Anderson Impurity Model}
\author{ B. Sriram Shastry and Edward Perepelitsky  }
\affiliation{ Physics Department, University of California, Santa Cruz, CA 95064, USA}
\author{Alex C. Hewson }
\affiliation{ Department of Mathematics, Imperial College, 180 Queen's Gate, London,   SW7 2BZ, United Kingdom}
\date{ \today}

\begin{abstract}

We apply the recently developed {\em extremely correlated Fermi liquid} theory     to the Anderson impurity model,
in the extreme correlation limit $U \rightarrow \infty$. We develop an expansion in a parameter  $\lambda$, 
related to  $n_d$, the average occupation  of the localized orbital,  and  find analytic expressions for the Green's functions to $O(\lambda^2)$. These yield the 
impurity spectral function  and also 
the self-energy  $\Sigma(\omega)$ in terms of the two self energies of the ECFL formalism.  The imaginary parts of the latter,  have roughly  symmetric low energy behaviour ($\propto
\omega^2$),  as predicted by Fermi Liquid theory. However, the inferred impurity 
self energy 
 $\Sigma''(\omega)$ develops    asymmetric corrections near $n_d
\to 1$, leading in turn to a strongly asymmetric impurity spectral function with a skew towards the occupied states.  Within this approximation the Friedel sum rule is satisfied  but we overestimate
the quasiparticle weight $z$ relative to the known exact results, 
resulting  in  an over broadening of the Kondo  peak.  Upon scaling the frequency  by the quasiparticle weight $z$,   the spectrum is found to be in reasonable agreement with numerical renormalization group results over a wide range of densities.

\end{abstract}
\maketitle



\section{Introduction and motivation}

The {\em Extremely Correlated Fermi Liquids} (ECFL)  theory  has been recently developed to understand the physics of correlations in the limit of infinite $U$- and applied to the \tJ model in \refdisp{ECFL} and in \refdisp{Monster}.  Here we apply the ECFL theory to the problem of the spin-$\frac{1}{2}$    Anderson impurity model (AIM) at $U = \infty$. The ECFL theory
 is based on a systematic expansion of the formally exact Schwinger equations of motion of the model for the (Gutzwiller) projected electrons in powers of  a parameter $\lambda$.  This parameter  is argued to be related to  $n$ the density of particles in the \tJ model, and in the same spirit, to  $n_d$ the average impurity level occupancy in the Anderson model considered here. Thus at low enough densities of particles, the complete  description of the system, including its dynamics  is expected   in  {\em quantitative} terms,  with  just a few terms in the $\lambda$ expansion.  Presently the theory to  $O(\lambda^2)$ has been evaluated for the \tJ model \refdisp{Monster}, and higher order calculations  in $\lambda$ valid up to higher densities  could be carried out in principle. We thus envisage systematically cranking up the density from the dilute limit, until we hit  singularities arising from phase transitions near $n\sim 1$   \cite{statmech}. This represents a possible road map for solving  one of the hard problems of condensed matter physics and is exciting for that reason.

  We apply the ECFL theory equations  to  $O(\lambda^2)$ to 
 the AIM  model in this work. This problem was introduced by Anderson
 \refdisp{pwa-aim} in 1961, and has been a fertile ground  where several fruitful ideas and  powerful techniques have been developed, and  tested against experiments in Kondo,   mixed valency and heavy Fermion  systems. These include  the  renormalization group ideas- from the intuitive poor man scaling of Anderson \cite{pwa-pms,haldane}, to the powerful numerical renormalization group (NRG) of Wilson \cite{w},  Krishnamurthy {\em et.al.} \cite{kww}, and more recent work in \cite{alt1,logan}.  A comprehensive review of the AIM and many popular  techniques used to study it, such as the large $N$ expansion \cite{largeN1,largeN2}, slave particles  \cite{slave} and the Bethe {\em ansatz} \cite{Bethe}
  can be  found in   \refdisp{hewson}.  In the AIM,  the Wilson renormalization group method provides  an essentially exact solution of the crossover from weak to strong coupling, without any intervening singularity in the coupling constant.  As emphasized in \cite{yamada,nozieres,rpt}, the ground state is asymptotically a Fermi liquid at all densities. This implies that  as a function of the density  $n_d$ (at any $U$),  the Fermi liquid  ground state evolves smoothly  without encountering any singularity,  from the low density limit (the empty orbital limit) to the intermediate density limit (the mixed valent regime), and finally   through to the  very high density limit (Kondo regime). In view of the non singular evolution in density, the AIM  provides us with  an ideal problem to  benchmark the basic ECFL ideas discussed above.
  
   The current understanding of the AIM model from \cite{kww,nozieres,yamada},  is that  Fermi liquid ground state and its attendant excitation spectrum are reached in the asymptotic sense, i.e. at low enough energies and T.  Our present  study of this model is somewhat broader.  We  wish to understand the excitations of the model in an enlarged region, in order to additionally  obtain an estimate  of the magnitude of  corrections to the asymptotic behaviour. To motivate this remark, note that  the ECFL formalism yields an asymmetry in the excitations and the spectral functions of the \tJ model for sufficiently high densities,  with a pronounced skew towards $\omega<0$, arising fundamentally  from Gutzwiller projection. This skew can   be interpreted as an asymmetric  correction to the leading particle-hole symmetric excitation spectrum of that model \refdisp{asymmetry} (e.g.   corrections to  $\Sigma''(\omega) \sim\{\omega^2+(\pi k_B T)^2\}$ behaviour of the Fermi liquid of the form  $\Sigma''(\omega)\sim\omega^3$). Such corrections    have been argued to be of central importance in explaining the anomalous lines shapes in the angle resolved photo emission spectra of High Tc superconductors in the normal state \refdisp{asymmetry} and \refdisp{gweon}.  Therefore it is  useful and  important to understand  the line shape and self-energy asymmetry   in  controlled calculations of  the Anderson model with infinite $U$, which shares the local Gutzwiller  constraint with the \tJ  model on a lattice.     A necessary condition for substantial asymmetry of the type seen in ECFL at $U = \infty$,  appears to be a large $U$, and hence is  difficult to  find  from a perturbative expansion in $U$ of the type pioneered   in \refdisp{yamada}. The study of the infinite $U$ limit of the AIM is therefore particularly interesting in the present context. AIM studies  of the  spectral functions \cite{Frota, Sakai, Costi-Hewson, Costi}  using NRG have become available in recent years.  We will compare our results with some of  these calculations later in this paper.

 In this paper, we use the ECFL machinery \refdisp{Monster} to obtain the exact Schwinger equation of motion for the d-electron Green's function and represent it in terms of two self-energies. These are further expanded in a series in the parameter $\lambda$ mentioned above, and the equations to second order are arrived at. These involve a second chemical potential $u_0$ that contributes to a shift in  the location of the localized energy level- bringing it closer to the chemical potential of the conduction electrons. The rationale for introducing this second chemical potential is similar to that in the \tJ model; the AIM possesses a shift invariance  identified in \disp{eq11}. Maintaining this invariance to different orders in $\lambda$ is possible only if we introduce $u_0$.  The second order equations are    studied numerically, and 
 the solution for the spectral function is compared with the NRG results. 

Since we expect some readers to be interested in the AIM more than  in the \tJ model, we  provide a fairly self-contained description of the ECFL method used here for the AIM.  In this spirit, it may be useful to  point out that  the  $\lambda$ parameter can be interpreted  by writing  a partially projected (d-orbital) Fermion operator $\hat{f}^\dagger_{ \si}(\lambda) = (1 - \lambda \ f^\dagger_{ \sib}f_{ \sib}) f^\dagger_{ \si}$ and its adjoint (here $\sib= -\si$). The operator $\hat{f}^\dagger_{ \si}(\lambda)$ interpolates between the unprojected Fermi operator   $f^\dagger_{ \si}$  at $\lambda=0$, and the Gutzwiller projected Hubbard operator $X^{\si 0}_i$ at $\lambda=1$.  The Hamiltonian is  written in terms of $\hat{f}^\dagger_{ \si}(\lambda), \ \hat{f}_{ \si}(\lambda)$, and expanding in $\lambda$ gives an effective Hamiltonian that generates the 
auxiliary Green's function $\GH$ below. As explained in   \refdisp{Monster},
the second (caparison) part also has an expansion in $\lambda$ that follows from the Schwinger equation and the product form \disp{GgmuAM}.
 
Below we first define our notations for  the model, and arrive at the exact Schwinger equation for the Green's function $\G$. Using a product ansatz $\G= \GH.\mu$, we obtain exact equations for the auxiliary Green's function $\GH$ and the caparison factor $\mu$. These are expanded in $\lambda$ and the second order equations are solved and compared with the NRG results for the spectral functions.

\section{ECFL Theory of Anderson Impurity  Model \lab{ECFLAM}}
\subsection{Model and Equations for the Green's Function}
We consider the Anderson impurity model  in the limit $U \rightarrow \infty$ given by  the following Hamiltonian:
\barray H &=& \sum_\sigma \epsilon_d X^{\sigma\sigma} +\sum_{k\sigma}{\epsilon}_kn_{k\sigma} \nn \\
&& +\factor \sum_{k\sigma}(V_k\ X^{\sigma0} \ c_{k\sigma}+V_k^* \ c^\dagger_{k\sigma} \ X^{0\sigma}),\earray
where $\Omega$ is the box volume, and  we have set the Fermi energy of the conduction electrons to  zero. Here $X^{a b}= |a\rangle \langle b|$ is the Hubbard projected electron operator with $|a\rangle$ describing  the empty orbital, and the two singly occupied   states $a=0, \pm \si$. 
 We  study the impurity Green's function:
\beq
\G_{\si_i \si_f}(\tau_i,\tau_f)= - \llangle   \ X^{0 \si_i}(\tau_i) \; X^{\si_f 0}(\tau_f) \label{Green}\rrangle,
\eeq
with $T_\tau$ the imaginary time ordering symbol,  the definition for an arbitrary time dependent operator $Q$: $\llangle Q \rrangle = \langle Tr \ T_\tau \ e^{- {\cal A}} Q \rangle /  \langle Tr \ T_\tau \ e^{- {\cal A}}  \rangle$, and
with the Schwinger source term ${\cal A} = \int_0^\beta \ d\tau \ \V^{\si_1 \si_2}(\tau) \ X^{\si_1 \si_2}(\tau) $, involving a Bosonic time dependent potential $\V$. 
Often we abbreviate $\V(\tau_i)\to \V_i$. As usual this potential is  set to zero at the end of the calculation.    In this paper  expressions such as $\G(\tau_i,\tau_f)$  and $\V$ are understood as $2\times2$ matrices in spin space.   
We assume a constant hybridization $V_k =V_0$, and a (flat) band of half-width $D$ with constant  density of states $\rho(\epsilon) = \rho_0 \ \theta(D- |\epsilon|)$ with 
 $\rho_0= \frac{1}{2D}$.

Taking the time derivative of \disp{Green} we obtain an equation of motion (EOM)
\barray
&&\{( \partial_{\tau_i} + \epsilon_d ) \iden + \V_i \}\G(\tau_i,\tau_f) = -\delta(\tau_i-\tau_f) \times (\iden - \gamma(\tau_i)) \nn \\
&&- \frac{1}{\sqrt{\Omega}} \left[ \iden- \gamma(\tau_i) + \D _i \right]. \sum_k V_k \ G(k, \tau_i;\tau_f), \label{eq3} 
\earray
where $\gamma(\tau_i)= \G^{(k)}(\tau_i^-,\tau_i)$ following \refdisp{ECFL} Eq.~(35), or more explicitly in terms of spin indices as  $\gamma_{\si_i \si_f}(\tau_i) = \si_i \si_f \G_{\sib_f \sib_i}(\tau_i,\tau_i^+)$, and with $\sib= -\si$
we introduced the mixed Green's function $G_{\si_i \si_f}(k, \tau_i; \tau_f) = -\llangle c_{k \si_i}(\tau_i) X^{\si_f 0}(\tau_f) \rrangle$, and a functional derivative operator $(\D_i)_{\si_i \si_j}=
(\si_i \si_j) \ {\delta }/{\delta \V^{\sib_i \sib_j}(\tau_i)}$. In the ECFL formalism \refdisp{ECFL},  \disp{eq3}  and similar equations are to be understood as matrix equations in spin space. Here the  higher order Green's functions  have been expressed in terms of the source functional  derivatives of the  basic ones; an   example  illustrates this:  $\si_i \si_j \llangle X^{\sib_i \sib_j} Q \rrangle = (\gamma_i - \D_i) \llangle Q \rrangle$. Proceeding further, we take a time derivative to find
\barray
(\partial_{\tau_i} + \epsilon_k) G(k, \tau_i;\tau_f) = -\factor V_k^* \ \G(\tau_i, \tau_f), \label{eq4}
\earray
so combining with \disp{eq3} we find the exact EOM
\barray
&&\{( \partial_{\tau_i} + \epsilon_d ) \iden + \V_i \}\G(\tau_i,\tau_f) = -\delta(\tau_i-\tau_f) \times (\iden - \gamma(\tau_i)) \nn \\
&&- \left( \iden- \gamma(\tau_i) + \D _i \right). \ \Delta(\tau_i-\tau_\bb{j}). \  \G( \tau_\bb{j};\tau_f), \label{eq5} 
\earray
with the convention that the  time label in bold letters $\tau_\bb{j}$ is to be integrated over $\in[0,\beta]$.  The   conduction   band enters through the ($\V$ independent) function
\beq \Delta(\tau_i-\tau_j) = -  \frac{\iden}{\Omega}  \sum_k |V_k|^2 (\partial_{\tau_i}+{\epsilon}_k)^{-1}\delta(\tau_i-\tau_j),  \eeq
with a  Fourier transform
\beq \Delta(i\omega_n) = \frac{1}{\Omega} \sum_k \frac{|V_k|^2}{i\omega_n-\epsilon_k} = V_0^2  \int \frac{\rho(\epsilon) \ d \epsilon}{i\omega_n-\epsilon}.  \label{deltafreq} \eeq
We will require below its  analytic continuation $i\omega_n\rightarrow \omega+i\eta$:
\barray
 &&\Delta(\omega+i\eta)=\Delta_R(\omega)-i \ \Gamma(\omega); \\
&&\Gamma(\omega)=\pi \ V_0^2 \ \rho(\omega) ; \;\;
\Delta_R(\omega)=  \frac{\Gamma_0}{\pi}\log\frac{|\omega+ D|}{|\omega- D|}.\label{DeltaAnCont}\nn  \\ 
\earray
Here $\Gamma_0 = {\pi} V_0^2 \rho_0$.
We now use the non-interacting Green's function
\beq \GHI_0(\tau_i,\tau_f) =  - (\partial_{\tau_i} +\epsilon_d + \V(\tau_i))\delta(\tau_i-\tau_f) - \ \Delta(\tau_i,\tau_f),  \label{g0}\eeq
and rewrite the fundamental equation of motion \disp{eq5} as
\beq \{\GHI_0(\tau_i,\tau_{\bb{j}})  + (\gamma_i - \D_i).\Delta(\tau_i-\tau_\bb{j} )\}. \G(\tau_\bb{j}, \tau_f) = (\iden - \gamma_i)\delta(\tau_i-\tau_f). \label{eq11} \eeq
\newline
Let us note an important {\em shift invariance} of \disp{eq11} and \disp{g0}. If we consider a transformation $\Delta(\tau) \to \Delta(\tau) + u_t \times \delta(\tau)$ with an arbitrary $u_t$, it is possible to show that \disp{eq11} is  unchanged, except for a shift  of $\epsilon_d$  by $-u_t$.    The added term $u_t \times (\gamma_i - \D_i). \G(\tau_i, \tau_f) $ vanishes upon using the Pauli principle and the Gutzwiller projection  applicable to operators {\em at the same time instant}. We use this  { shift invariance}  below, to introduce a second chemical potential.   
In  the ECFL theory, we use a product {\em ansatz}
\beq \G(\tau_i,\tau_f) = \GH(\tau_i,\tau_{\bb{j}}) \ . \ \mu(\tau_{\bb{j}},\tau_f) \label{GgmuAM}\eeq
where $\mu$ is  the caparison factor, and use this in \disp{eq11}. It is useful to introduce two  vertex functions 
$
\Lambda^{\si_1 \si_2}_{\si_3 \si_4}(\tau_n,\tau_m;\tau_i)  = - \frac{\delta}{\delta \V^{\si_3 \si_4}_i} \GHI_{\si_1 \si_2}(\tau_n,\tau_m), $ and $
\U^{\si_1 \si_2}_{\si_3 \si_4}(\tau_n,\tau_m;\tau_i)  =   \frac{\delta}{\delta \V^{\si_3 \si_4}_i} \mu_{\si_1 \si_2}(\tau_n,\tau_m)$ as usual, and 
suppressing the time indices, we note $\frac{\delta}{\delta \V} . \GH= \GH. \Lambda.\GH $. We now use the  chain rule  and \disp{GgmuAM} to write
 $\D.\Delta.\G=\D. \Delta. \GH.\mu= \xi^*. \Delta. \GH.\Lambda_*.\GH. \mu+ \xi^*.\Delta. \GH.\U_*$, with the matrix $\xi_{\si \si'} = \si \si'$.  The  $*$  symbol from \refdisp{ECFL} is illustrated in component form by an example:  $\cdots \xi^*_{\si_a \si_b} \cdots\delta/\delta \V^{*} = \cdots \si_a \si_b \cdots \delta/\delta \V^{\sib_a \sib_b}$, or in terms of the vertex functions  $\cdots \xi^*_{\si_a \si_b} \cdots \Lambda^{\si' \si''}_*\cdots  = \cdots \si_a \si_b  \cdots \Lambda^{\si' \si''}_{\sib_a \sib_b}\cdots$, with the upper indices of $\Lambda$ governed by the rules of the  matrix product.
 Following \refdisp{ECFL} we define the linear operator $\LL(i,j) = \xi^*.\Delta(i,\bb{j}).\GH(\bb{j},j).\frac{\delta}{\delta \V_i^*}$. We can now collect these definitions to rewrite
 $\D.\Delta.\G= \xi^*.\Delta. \GH. \Lambda_*.\GH. \mu + \xi^*.\Delta. \GH.\U_* = \Phi.\GH. \mu + \Psi$,
 and define the two self-energies:
 \barray
 \Phi(i,j)&=& -\LL(i,\bb{r}).  \GHI(\bb{r},j)=\xi^*.\Delta(i,\bb{j}).\GH(\bb{j},\bb{k}).\Lambda_*(\bb{k},j;i);  \nn \\
 \Psi(i,j)&=& \LL(i,\bb{r}).  \mu(\bb{r},j)=\xi^*.\Delta(i,\bb{j}).\GH(\bb{j},\bb{k}).\U_*(\bb{k},j;i).  \nn \\ \label{phipsi}
 \earray
  Summarizing, we may rewrite the exact EOM \disp{eq11}  symbolically:
 \barray
\{\GHI_0 + \gamma.\Delta - \Phi \}. \GH . \mu 
= (\iden - \gamma)\delta    +  \Psi
. \label{eq14} 
 \earray
 This equation is   split into two parts by requiring $\GH$ to be canonical:
\beq
 \GHI= \{\GHI_0 + \gamma.\Delta - \Phi \},\; \mbox{and}\;\; \mu 
= (\iden - \gamma)\delta    +  \Psi, \label{eq15}
\eeq
 bringing it into 
 the standard form in the  ECFL theory \refdisp{ECFL}.
  Using  \disp{phipsi},  we note that the formal solutions of \disp{eq15}  are: $\GHI= (\iden- \LL)^{-1} .\left( \GHI_0 +   \gamma . \Delta \right)$ and 
$\mu=   (\iden-  \LL)^{-1} .\left(\iden-    \gamma  \right)\delta$. We introduce the resolvent kernel $\L$ using
the identity $ (\iden-  \LL)^{-1}= \iden+  \L$ where $\L=  (\iden-  \LL)^{-1}.\LL$. In terms of the resolvent, we see that
 \beq
 \Phi = \L. (-\GHI_0 -  \gamma.\Delta), \; \mbox{and} \;
 \Psi= -  \L.  \gamma.\delta. \label{eq16}
 \eeq 
 Therefore distributing the action of $\L$ over the two terms, we can rewrite
 \barray
 \Phi&=& \chi+ \Psi.\Delta,  \label{eq17} \\
 \mbox{with} \;\chi &=& \L.(-\GHI_0).   \label{eq18}
 \earray
 Therefore  the self-energy $\Phi$ breaks up into two parts, as in \disp{eq17}.
 Note that in \disp{eq16}, the expressions $\gamma.\Delta$ and $\gamma.\delta$ involve multiplication at equal times, whereas in \disp{eq17}, $\Psi.\Delta$ implies a convolution in time.
 The two Green's functions satisfy the pair of sum rules
 \barray
 \GH(\tau,\tau^{+}) = \frac{n_d}{2};\;\;\;
\G(\tau,\tau^{+}) = \frac{n_d}{2},\;\;\;\label{sumrules1}
\earray  
where $n_d$ is the number of electrons on the d-orbital $n_d= \sum_\si \langle X^{\si \si} \rangle$.

  In the context of the \tJ model in \refdisp{Monster}, the sum rule for $\GH$ is necessary to satisfy the Luttinger-Ward theorem.  If we use the representation  $\hat{f}^\dagger_{ \si}(\lambda) = (1 -  \ f^\dagger_{ \sib}f_{ \sib}) f^\dagger_{ \si}$ for the correlated electrons,   this constraint  is understandable as the constraint on the number of ``uncorrelated'' Fermions $ \langle f^\dagger_\sigma f_\sigma \rangle$, which must agree with the number of  physical (correlated) electrons $\langle \hat{f}^\dagger_\sigma \hat{f}_\sigma \rangle$. Similarly, in the present case, this constraint is needed to fulfill the Friedel sum rule. We also remark 
  that the self-energy $\Psi$, unlike its counterpart $\Phi$,  is dimensionless, and thus  interpreted 
as an adaptive spectral weight  \cite{Monster}.

 \subsection{Zero Source Limit \lab{Zerosource}}
Upon turning off the sources, all objects become functions of only $\tau_i-\tau_f$ and may therefore be Fourier transformed to Matsubara frequency space. By Fourier transforming \disp{GgmuAM},  \disp{eq15} and \disp{eq17} and using $\gamma\to \frac{n_d}{2}$ we obtain the following expressions in frequency space:
\barray \G(i\omega_n) &=& \GH(i\omega_n) \ . \ \mu(i\omega_n), \nn \\
 \mu(i\omega_n) &=& 1- \frac{n_d}{2} \ + \  \Psi(i\omega_n), \nn \\
 \GHI(i\omega_n) &=&  i\omega_n -   \epsilon_d   -   \Delta(i\omega_n) \mu(i \omega_n) -   \chi(i\omega_n). \label{ecflG}
\earray
Alternately this result can be rewritten in terms of the Dyson-Mori self-energy representation as
\barray
\G(i \omega_n) &=& \frac{1- \frac{n_d}{2}}{i \omega_n - \epsilon_d - (1- \frac{n_d}{2}) \Delta(i \omega_n) - \Sigma_{DM}(i \omega_n)} \; \ \ \label{DMG}\;
\earray
and
\barray
&&\Sigma_{DM}(i \omega_n)+ \epsilon_d - i \omega_n = \nn \\
&& \frac{1- \frac{n_d}{2}}{1- \frac{n_d}{2}+ \Psi(i \omega_n)} \left( \chi(i \omega_n) + \epsilon_d - i \omega_n\right). \label{DMSigma}
\earray
 The sum rules \disp{sumrules1} are:
\barray
\sum_{i\omega_n} \G(i\omega_n)e^{i\omega_n\eta} = \frac{n_d}{2};\;\;\;
\sum_{i\omega_n} \GH(i\omega_n)e^{i\omega_n\eta} = \frac{n_d}{2}.\label{sumrules}
\earray
 We observe that the usual Dysonian self-energy $ \Sigma_{AM}( i\omega_n)$ defined through the usual Dyson equation (valid for finite $U$) $G^{-1}= i \omega_n - \epsilon_d - \Delta(i\omega_n)-\Sigma_{AM}( i\omega_n)$ in the infinite $U$ limit can be obtained from
\beq
\Sigma_{AM}(i \omega_n)= \frac{2}{2- {n_d}} \Sigma_{DM}(i \omega_n) + \frac{{n_d}}{2- {n_d}} (\epsilon_d - i \omega_n). \label{eq24}
\eeq
The unlimited growth with $ \omega_n$ makes  this self-energy somewhat inconvenient to deal with, and therefore motivated the introduction of the Dyson Mori object, which is better behaved in this regard. After analytic continuation $i \omega_n \to \omega+ i 0^+$,  the imaginary part  of $\Sigma_{AM}$ is well behaved and  finite as $\omega \to \infty$. It   is obtained from the NRG method and compared with the relevant ECFL functions after scaling by $1- \frac{n_d}{2}$ as in \disp{eq24}. 
 We notice that the density $n_d$ appears explicitly in the expressions for the Green's functions, and must therefore be calculated self-consistently, from \disp{sumrules}. 
This   feature is quite natural in the present approach, since \disp{eq3} for the Green's function contains $\gamma$ and therefore $n_d$ explicitly.   
 \subsection{Introducing $\lambda$ and $u_0$ into the equations.}
 Summarizing the work so far: \disp{eq15}, \disp{eq16} and \disp{eq17}  follow from \disp{eq11} upon using the product ansatz \disp{GgmuAM}, and  are  exact equations. In order to get concrete results, we proceed by introducing two parameters into the equations. 
 (I) The  parameter $\lambda\in[0,1]$  multiplies certain terms shown in \disp{2paras}, allowing a density type expansion, and continuously connects the uncorrelated Fermi system $\lambda=0$ to the extremely correlated case $\lambda=1$.  (II) The second parameter $u_0$ is introduced as shown in \disp{2paras}. It is the  second chemical potential  used  to enforce the shift identities of the exact equation \disp{eq11}. 
  \disp{eq11} now   becomes
 \beq \{ \GHI_0 + \lambda(\gamma-\D). (\Delta- \frac{u_0}{2}\delta)\}.\G= (\iden - \lambda \gamma) \delta. \label{2paras}\eeq
 As a consequence,  in \disp{eq14} to   \disp{eq18} we set $\gamma \to \lambda \gamma$, $\Psi \to \lambda \Psi$, and $\Phi \to \lambda \Phi$, or $\chi \to \lambda \chi$.   Secondly  in \disp{eq14} to   \disp{eq18} we set $\Delta(\tau_i,\tau_f) \rightarrow \Delta(\tau_i,\tau_f) - \frac{u_0}{2} \  \delta(\tau_i-\tau_f)$. Note that there is no  shift of  \disp{g0} implied in \disp{2paras}.

 We write \disp{eq15} with $\lambda$ inserted explicitly and the understanding  that $\Delta(\tau_i,\tau_f)$ has been shifted as (\refdisp{fny1}):
\barray
\GHI(\tau_i,\tau_f) &=& \GHI_0(\tau_i,\tau_f) + \ \lambda \gamma(\tau_i). \Delta(\tau_i,\tau_f) - \lambda \ \Phi(\tau_i,\tau_f)\label{ginAM}, \nn \\ 
 \mu(\tau_i,\tau_f) &=&  \delta(\tau_i-\tau_f)(\iden- \lambda\gamma(\tau_i)) + \lambda \ \Psi(\tau_i,\tau_f)\label{muAM}, \label{eq21}
 \earray
 where the two self-energies are given in terms of the vertex functions as
 \barray
  \Phi(\tau_i,\tau_f) & =&  \xi^*  .   \Delta(\tau_i,\tau_{\bb{j}}) .  \GH(\tau_{\bb{j}},\tau_{\bb{k}})  .  \Lambda_*(\tau_{\bb{k}},\tau_f;\tau_i) \nn \\
 \Psi(\tau_i,\tau_f)& =& \xi^* .  \Delta(\tau_i,\tau_{\bb{j}})  .  \GH(\tau_{\bb{j}},\tau_{\bb{k}}) .  \U_*(\tau_{\bb{k}},\tau_f;\tau_i). 
 \earray
 On switching off the sources, these expressions can be spin resolved and expressed as
 $  \Phi  =  \Delta \ \GH \ \Lambda^{(a)} $ and $  \Psi  =  \Delta \ \GH \ \U^{(a)} $, with the same time labels as above, and with the usual  spin decomposition $\Lambda^{(a)} =\Lambda^{\si \si}_{\sib \sib}- \Lambda^{\si \sib}_{\si \sib} $.
\subsection{$\lambda$ Expansion \lab{lamexp}} 
We note that we can obtain the equations of motion for the Anderson model from the infinite-d equations of motion for the $\tJ$ model by making the following substitutions and replacing all space-time variables with just time\cite{ECFLlarged}.
 \beq
t[i,f]\to -\Delta(\tau_i,\tau_f); \; \varepsilon_k \to \Delta(i \omega_k),\;J \to 0, \;
\chem \to -\epsilon_d. 
 \label{substitutions}
\eeq
The $\lambda$ expansion for the Anderson model is therefore analogous to the one  for the \tJ model in \refdisp{Monster} and the large-d \tJ model in \refdisp{ECFLlarged}, and   can be obtained from them by making the substitutions in \disp{substitutions} and changing all frequency momentum four vectors to  just frequency.  For completeness, Appendix A provides a brief derivation (in time domain) of the following equations.
 Denoting 
 \beq
 \sw= 1 - \lambda \frac{n_d}{2}+ \lambda^2 \frac{n_d^2}{4},
 \eeq
and  the  frequently occurring object  $${\cal R} =\GH(i\omega_p)\GH(i\omega_q)\GH(i\omega_p+i\omega_q-i\omega_n),$$   we obtain to $O(\lambda^2)$   the expressions :
\barray
\G(i \omega_n)& =& \GH(i \omega_n) \mu(i \omega_n), 
\mu(i\omega_n) = \sw+\lambda\Psi(i\omega_n),\; \; \;\;\label{eq31} \\
\GHI( i\omega_n) &=& i\omega_n - \epsilon_d' - (\Delta(i\omega_n) - \frac{u_0}{2})\mu(i \omega_n) \nn \\
&&
 -\lambda {\chi}(i\omega_n),\;\;\;\;\;\; \label{eq32} \\
 {\chi}(i\omega_n) &=& -\lambda\sum_{p,q}[2\Delta(i\omega_p)-u_0] \nn \\
&&\times [\Delta(i\omega_p+i\omega_q-i\omega_n)-\frac{u_0}{2}]{\cal R},\label{eq33} \\
 {\Psi}(i\omega_n) &=& -\lambda\sum_{p,q}[2\Delta(i\omega_p)-u_0]{\cal R}. \label{glambda}
 \earray
 The energy $\epsilon_d'$ is given by collecting the static terms in $\Phi$ as
\beq \epsilon_d' = \epsilon_d + u_0(\lambda\frac{n_d}{2}-\lambda^2\frac{n_d^2}{8})
+ \frac{u_0}{2} \sw
-\lambda\sum_{i\omega_p}\Delta(i\omega_p)\GH(i\omega_p). \label{chem} \eeq
The shift-theorem is satisfied  by all the terms separately- since we have taken care to form expressions of the type $\Delta- \frac{u_0}{2}$.  As discussed  in \refdisp{Monster}, the shift theorems mandate the introduction of $u_0$, and its availability,  in addition to $\epsilon_d$, enables us to  fix the pair of  sum rules  \disp{sumrules1}. As explained, we must set $\lambda \to 1$ before using these expressions. 

 Within the $O(\lambda^2)$ theory, the total spectral weight of the Green's function is $\sw$ rather than the exact value $1-\frac{n_d}{2}$. This is understood as the incomplete projection to singly occupancy leading to an excess in the total number of states available to the system. In order to ensure that $\Sigma_{DM}(\omega)$ retain the feature of being finite as $\omega \to \infty$, it must be slightly redefined (to $\hat{\Sigma}_{DM}$) in the  $O(\lambda^2)$ theory.
\beq G(\omega) = \frac{\sw}{\omega - \epsilon_d''-\sw\Delta(\omega)-\hat{\Sigma}_{DM}(\omega)} \label {DMGhat}\eeq
where 
\beq \epsilon_d'' \equiv \epsilon_d' -\frac{u_0}{2}\sw \label{ed''}\eeq
Using \disp{eq30} and \disp{eq32}, we can relate $\hat{\Sigma}_{DM}(\omega)$ to $\chi(\omega)$ and $\Psi(\omega)$.
\beq \hat{\Sigma}_{DM}(\omega) + \epsilon_d' - \omega = \frac{\sw}{\sw+\Psi(\omega)}(\chi(\omega)+\epsilon_d'-\omega)\label{DMhatpsichi}\eeq
Since $\Psi(\omega)$,$\chi(\omega)\to0$ as $\omega \to \infty$, we see explicitly that $\hat{\Sigma}_{DM}(\omega) $ remains finite in this limit. Just as in the case of $\Im m  \ \Sigma_{DM}(\omega)$, $\Im m \  \hat{\Sigma}_{DM}(\omega)$ is related to 
$\Im m \ \Sigma_{AM}(\omega)$ by a multiplicative constant ($1-\frac{n_d}{2}$ and $\sw$ respectively), and therefore their spectra are identical apart from this multiplicative constant. Comparing \disp{DMG} and \disp{DMGhat}, we see that the latter is obtained from the former with the substitutions
\beq
\Sigma_{DM}(\omega) \to \hat{\Sigma}_{DM}(\omega); \; \epsilon_d \to \epsilon_d'';\;
1-\frac{n_d}{2} \to \sw. 
 \label{subsDM}   
 \eeq
 Keeping these substitutions in mind, we will now only use $\Sigma_{DM}(\omega)$ from the exact theory, with the understanding that the same expressions hold for $\hat{\Sigma}_{DM}(\omega)$ in the $O(\lambda^2)$ theory as long as the substitutions in \disp{subsDM} are made.
 
 \subsection{Friedel Sum Rule at $T=0$}
At $T=0$, the Friedel sum rule \cite{friedel,langer,langreth} plays an important role in the AIM, parallel to that of the Luttinger-Ward volume theorem in Fermi liquids. In \refdisp{langreth}, the original form of the  Friedel sum rule is written in terms of $\eta_\si(\omega)$,  the phase shift  of the conduction electron with spin $\si$ at energy $\omega$:
\barray
\eta_\si(\omega)& =& \frac{1}{2 i} \log \left[ \G_\si(\omega+ i 0^+)\G_\si^{-1}(\omega- i 0^+) \right] , \label{usefulfsr}
\earray
where the logarithm is chosen with a branch cut along the positive real axis, so that $0 \leq \eta \leq \pi$.
The Friedel sum rule is then written as :
\beq \eta_\si(\omega=0)= \frac{\pi n_d}{2}. \label{fsr-original} \eeq
This theorem is proven for the AIM at  finite $U$  \refdisp{langreth}, by adapting  the  argument of Luttinger and Ward \refdisp{lw}, with an implicit assumption   of a non-singular evolution in $U$ from 0. We assume that the Friedel sum rule also holds in the extreme correlation limit $U \rightarrow \infty.$ Using  the Dyson Mori representation \disp{DMG} to compute the phase shift in \disp{usefulfsr}, we may  rewrite  this  as
\beq
{n_d}=1- \frac{2}{\pi} \tan^{-1} \left[ \frac{ \epsilon_d + \Re e \Sigma_{DM}(0)}{\Gamma_0 ( 1- \frac{n_d}{2})}\right], \label{fsr-2}
\eeq
with $\epsilon_d +\Re e \Sigma_{DM}(0)>0$, in  the physical case of $0 \leq n_d \leq 1$.
 It is easily seen \cite{fn-2} that this form is equivalent to the standard statement of the Friedel sum rule(\refdisp{hewson}):
\beq
\rho_{\G}(0) = \frac{1}{\pi \Gamma_0} \ \sin^2(  \frac{ \pi n_d}{2}) , \label{fsr-1}
\eeq
Within the approximation of the $\lambda$ expansion, the Friedel sum rule implies a relationship between the values of the two self-energies at zero frequency. 
\beq
 n_d =1- \frac{2}{\pi} \tan^{-1} \left[ \frac{\epsilon'_d- \frac{u_0}{2} \mu(0) + \chi(0) }{\Gamma_0 \mu(0) }\right], \label{fsr-4}
\eeq
This can be obtained by using the substitutions from \disp{subsDM} in \disp{fsr-2}, and using Eqs. (\ref{DMhatpsichi}),(\ref{ed''}), and (\ref{eq30}).

We can also record  a  result for the auxiliary density of states $\rho_{\GH}(\omega=0) $, analogous to \disp{fsr-1} here. It  follows from \disp{rhog}, with the Fermi liquid type assumption of vanishing of $\rho_\Psi(0)$ at $T=0$, and reads
\beq
\rho_{\GH}(0) = \frac{1}{\pi \Gamma_0 \mu(0)} \ \sin^2(  \frac{ \pi n_d}{2})  \label{fsr-5}
\eeq
We check the validity of the Friedel sum rule within the $\lambda$ expansion in both the forms \disp{fsr-1} and \disp{fsr-4}.
In doing so, we are thus testing if the strategy of  the two ECFL  sum rules \disp{sumrules} enforces the Friedel sum rule, in a  situation  that is essentially different from that in  finite $U$ theories so that the central result of Luttinger and Ward \refdisp{lw} is not applicable in any obvious way.

\subsection{Computation of Spectral function \lab{spectral}}

In computing the spectral function, we follow the approach taken in \refdisp{Monster}, in which the spectral function is calculated for the $O(\lambda^2)$ ECFL theory of the $\tJ$ model. Our calculation is made simpler due to the absence of any spatial degrees of freedom, but more complicated by the presence of the frequency dependent factor $\Delta(i\omega_n)$. We define the various spectral functions and the relationships between them. These expressions are analogous to those in sec.$\MakeUppercase{\romannumeral 3}$ A of \refdisp{Monster}.

\beq Q(i\omega_n) = \int_{-\infty}^{\infty} d\nu \frac{\rho_Q(\nu) }{i\omega_n - \nu} \eeq
Where $Q$ can stand for any object such as $\G$, $\GH$, ${\chi}$, $\Sigma_{DM}$ or $\Psi$. Therefore after analytic continuation $i\omega_n \rightarrow \omega+i 0^+$
\barray \rho_Q(\omega) \equiv -\frac{\Im m}{\pi}    Q(\omega+i 0^+) \ \mbox{and} \;
 \Re e \ Q(\omega)  = \mathcal{H}[\rho_Q](\omega), \nn \\ \label{hilbert} \earray
 where for any real density $\rho_Q(\omega)$  the Hilbert transform is denoted as
$ \mathcal{H}[\rho_Q](\omega) = P \int_{-\infty}^{\infty} d\nu \frac{\rho_Q(\nu)}{\omega-\nu} $.
From \disp{glambda}, we find that
\beq \rho_\G(\omega) = \rho_{\GH}(\omega)[\sw +\Re e \ \Psi(\omega)] + \rho_\Psi(\omega)\Re e \ \GH(\omega) \label{rhog} \eeq
With $f(\omega)=\frac{1}{1+e^{\beta\omega}}$ and $\bar{f}(\omega) = 1 - f(\omega)$, the two sum rules \disp{sumrules} read  
\beq \int_{-\infty}^{\infty} f(\omega) \ \rho_{\GH}(\omega) \ d\omega = \frac{n_d}{2} \ , \ \int_{-\infty}^{\infty} f(\omega) \ \rho_{\G}(\omega) \ d\omega = \frac{n_d}{2} .\eeq
  We also note $\rho_{\Delta}(\omega)=\frac{\Gamma(\omega)}{\pi}$. It is useful to define a mixed (composite) density
  \beq
  \rho_M(x)= \rho_\GH(x) (\Delta_R(x) - \frac{u_0}{2}) + \rho_\Delta(x) \Re e \ \GH(x), \label{composite}
  \eeq
  so that we can integrate (or sum) the internal frequencies in  \disp{glambda} efficiently (see Appendix B), and write the two relevant complex self-energies (with $\omega \equiv \omega+ i 0^+$) as 
\barray
\Psi(\omega) &=& - 2 \lambda \int_{u, v, w} \frac{\rho_M(u) \rho_\GH(v)\rho_\GH(w)}{\omega - u - v+w}\nn \\
&&\times  \left[ f(u) f(v) \bar{f}(w) + \bar{f}(u) \bar{f}(v) {f}(w)\right]\nn \\
\chi(\omega) &=& - 2 \lambda \int_{u, v, w} \frac{\rho_M(u) \rho_\GH(v)\rho_M(w)}{\omega - u - v+w}\nn \\
&&\times  \left[ f(u) f(v) \bar{f}(w) + \bar{f}(u) \bar{f}(v) {f}(w)\right]\label{selfenergies}
\earray
In these expressions $u,v,w$ are understood to be real variables, and using \disp{hilbert} we can extract the real and imaginary parts of $\Psi$ and $\chi$ in terms of the spectral functions.
  

\begin{table}[b] \label{table1}
    \begin{tabular} {| p{.68cm}  | p{1.8cm}  | p{1.4cm}  | p{1.2cm} |p{1.1cm} |p{1.1cm} |}
    \hline
    $n_d$ & $\rho_{G,ECFL}(0)$ & $\epsilon_{d,ECFL}$ & $\epsilon_{d,NRG}$ & $z_{ECFL}$&$z_{NRG}$ \\ \hline
    0.35& 8.69001 + 1.80298 \%&-0.00326&-0.00328& 0.75278& 0.69676  \\ \hline
    0.441& 12.9824 + 1.1388 \%&-0.00958 &-0.0094  &0.66073& 0.56704\\ \hline
    0.536& 17.7117 + 0.72518\%&-0.01518 &-0.01473 & 0.55883& 0.41649\\ \hline
    0.6& 20.8337 + 0.40918 \% &-0.018870&-0.01800 & 0.48934& 0.31249\\ \hline
    0.7& 25.2704 + 0.62054\%& -0.02387&-0.02387& 0.38807& 0.16912\\ \hline
   0.777& 28.0824 + 0.25626\%&-0.03147&-0.02947&0.31380 &0.08065\\ \hline
   0.834 &29.7154 + 0.20342 \% & -0.03744&-0.03519 &0.26484&0.03510\\
    \hline
    \end{tabular}
    \caption{The bare impurity level $\epsilon_d$ as well as the quasiparticle weight $z$ are displayed for the ECFL and the NRG calculations for all values of the density. Additionally, the theoretical value for the Friedel sum rule as well as the ECFL deviation from it are displayed. }
    \end{table}

\section{Results}
We calculated the spectral functions $\rho_G$ , $\rho_\Sigma$ , $\rho_\chi$ , and $\rho_\Psi$ using the values $D=1$, $\Gamma_0=0.01$, and $T=0$. The zero temperature limit is easily achieved in the ECFL theory by setting all of the Fermi functions to step functions. We expect that the spectral function calculated within the ECFL $O(\lambda^2)$ theory will be accurate through a density of approximately $n=0.6$.  The source of this error estimate is the high frequency behaviour within the $\lambda$ expansion  of the Green's function \disp{glambda}  $\G\sim \frac{\sw}{i \omega}$, this deviates from the known exact behaviour $\G\sim \frac{1- {n_d}/{2}}{i \omega}$.
 The error  grows with increasing density, but we expect to have reasonable results even at $n=0.7$.

\begin{figure}[h]
\begin{center}
\includegraphics[width=.45\columnwidth] {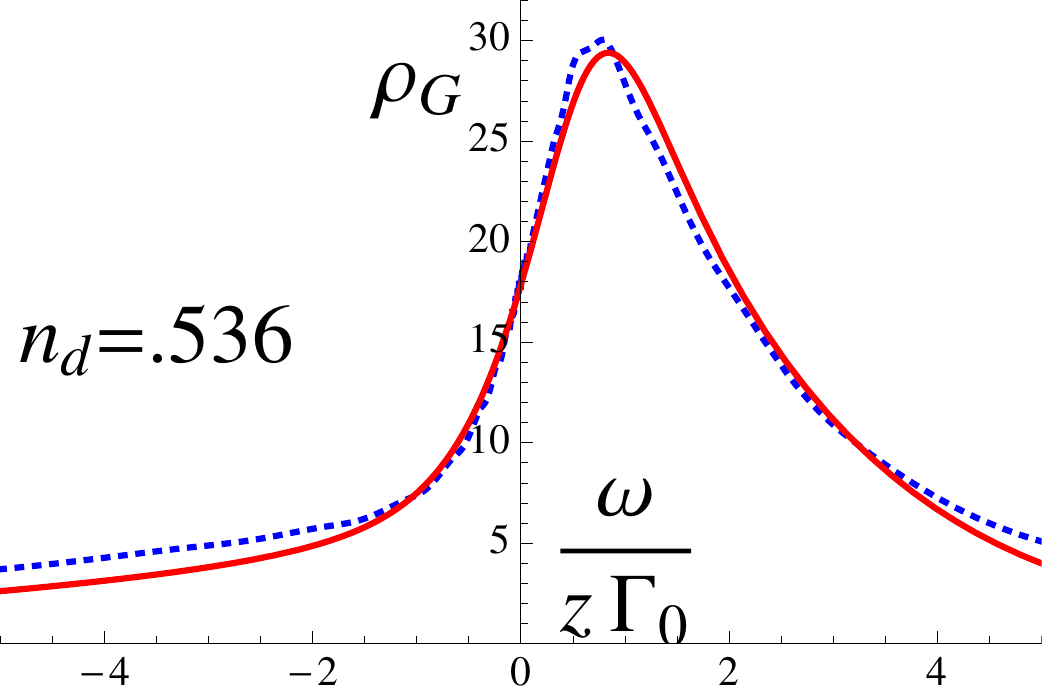}
\includegraphics[width=.45\columnwidth] {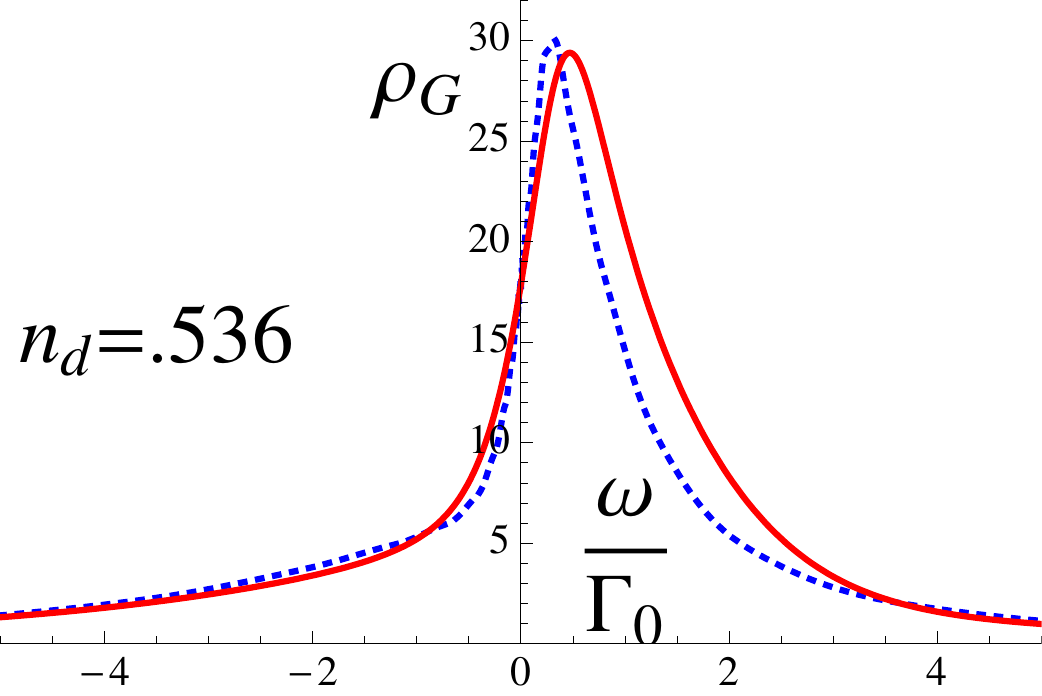}
\end{center}
\caption{The spectral density for the physical Green's function for the density of $n_d=0.536$. For the plot on the left, both the ECFL and NRG curves are plotted versus $\frac{\omega}{\Gamma_0 z}$. Since ECFL has a larger z value, the absolute scale of the $\omega$ axis differs for the two curves. For the plot on the right, both ECFL and NRG are plotted versus $\frac{\omega}{\Gamma_0 }$ and hence the ECFL peak is too wide. }
\label{rhoGscale}
\end{figure}

\begin{figure}[h]
\begin{center}
\includegraphics[width=\columnwidth] {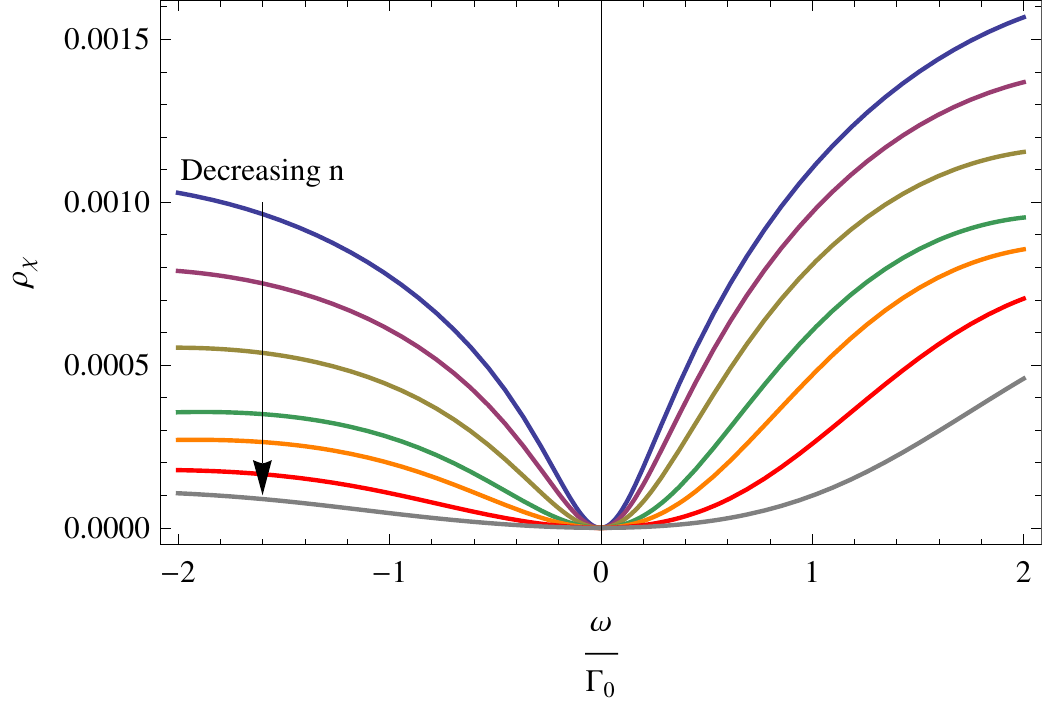}
\end{center}
\caption{The spectral function for $\chi$ for densities of $n_d=0.834, 0.777,
  0.7, 0.6, 0.536, 0.441, 0.35.$ }
\label{rhoChibeforel}
\end{figure}

\begin{figure}[h]
\begin{center}
\includegraphics[width=\columnwidth] {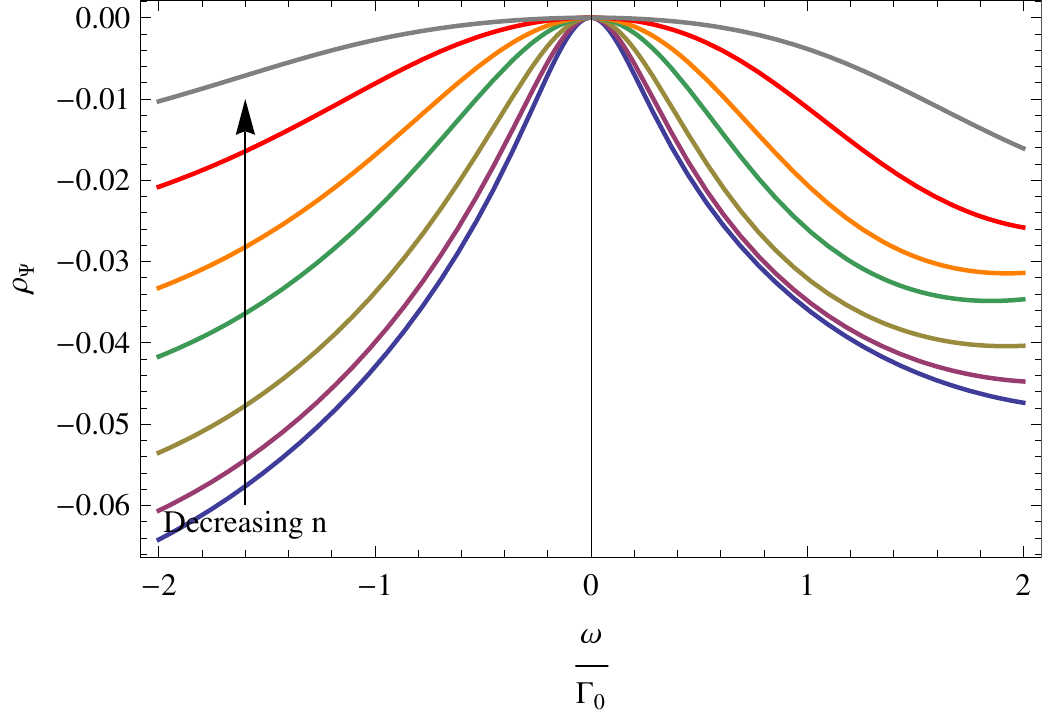}
\end{center}
\caption{The spectral function for $\Psi$ for densities of $n_d=0.834, 0.777,
  0.7, 0.6, 0.536, 0.441, 0.35.$ }
\label{rhoPsibeforel}
\end{figure}

In Table~(I), we show the results for the spectral function at zero energy in
terms of the percentage deviation from the Friedel sum rule \disp{fsr-1},
demonstrating that the ECFL satisfies the Friedel sum rule to a high degree of
accuracy. We specify the occupation number $n_d$ and show the
values of the  energy level $\epsilon_d$ and quasiparticle weight $z$ calculated within the ECFL and NRG calculations. The values of $\epsilon_d$ are in good agreement between the two calculations, while there is a discrepancy in $z$ which becomes more pronounced at higher densities. Its significance is discussed below.
\begin{widetext}   
 In \figdisp{rhoGbefore} we display the spectral functions at the indicated densities- indicating a smooth evolution with density. The Kondo  or Abrikosov-Suhl resonance at positive frequencies becomes sharper as we increase density and moves closer to $\omega =0$. If the ECFL and NRG spectral functions are compared (as in right panel of \figdisp{rhoGscale} for $n_d=0.536$), one will find that the peak in the ECFL spectral function is over broadened. This over broadening becomes worse at larger densities and better at lower densities. However, it can be understood well in terms of the elevated value of $z$ for ECFL at higher densities. Hence, before doing the comparison, we first rescale the $\omega$ axis for both the ECFL and NRG spectral functions by the appropriate $z$ (as in the left panel of \figdisp{rhoGscale} for $n_d=0.536$ and in \figdisp{rhoGbefore} for the other densities). 
   They are then found to be in good agreement. We also found good agreement with the NRG spectral functions in \refdisp{Costi}. The ECFL spectral function $\rho_G$ is constructed out of the two spectral functions $\rho_\chi$ and $\rho_\Psi$ that are shown at various densities in \figdisp{rhoChibeforel} and \figdisp{rhoPsibeforel}, exhibiting Fermi liquid type quadratic frequency dependence at low $\omega$.
     \begin{figure}
   \begin{center}
\includegraphics[width=.3\columnwidth] {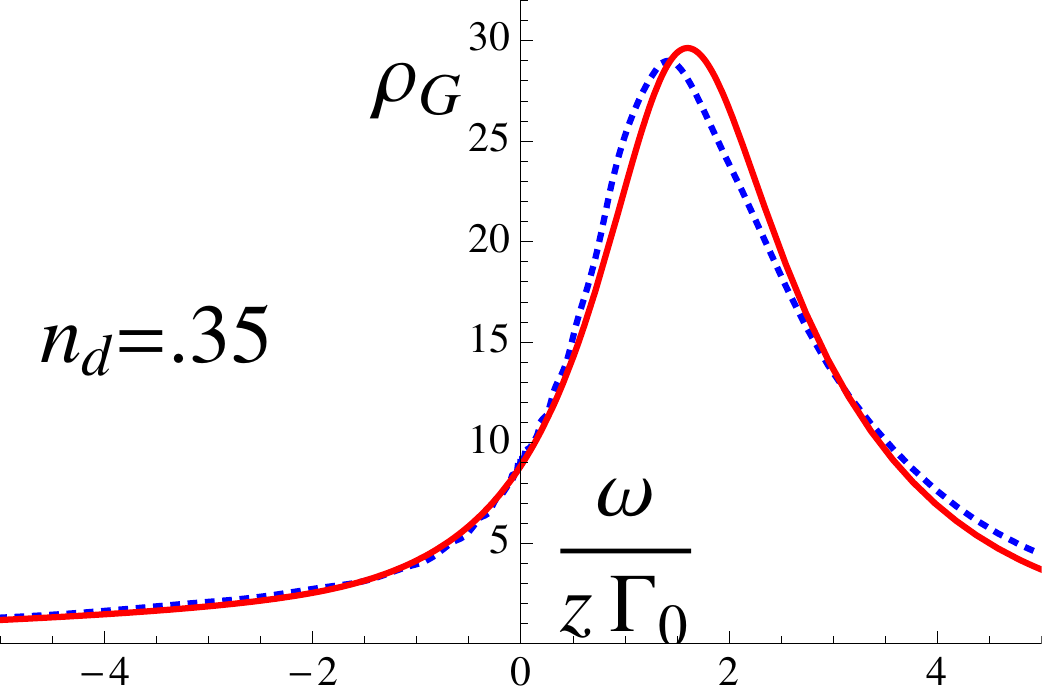}
\includegraphics[width=.3\columnwidth] {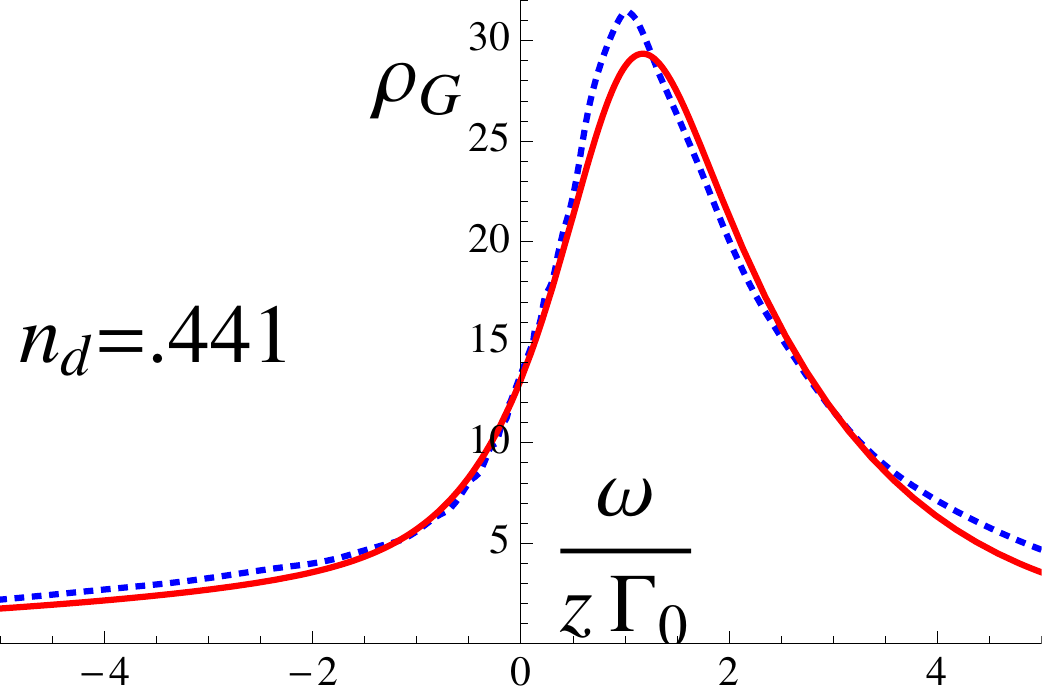}
\includegraphics[width=.3\columnwidth] {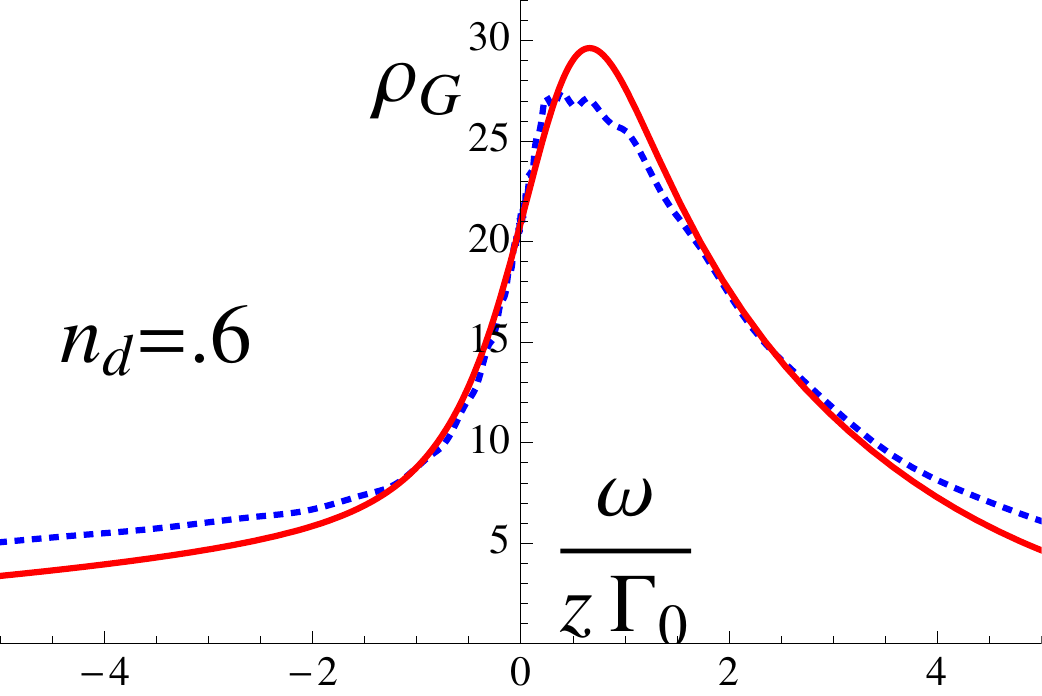}
\includegraphics[width=.3\columnwidth] {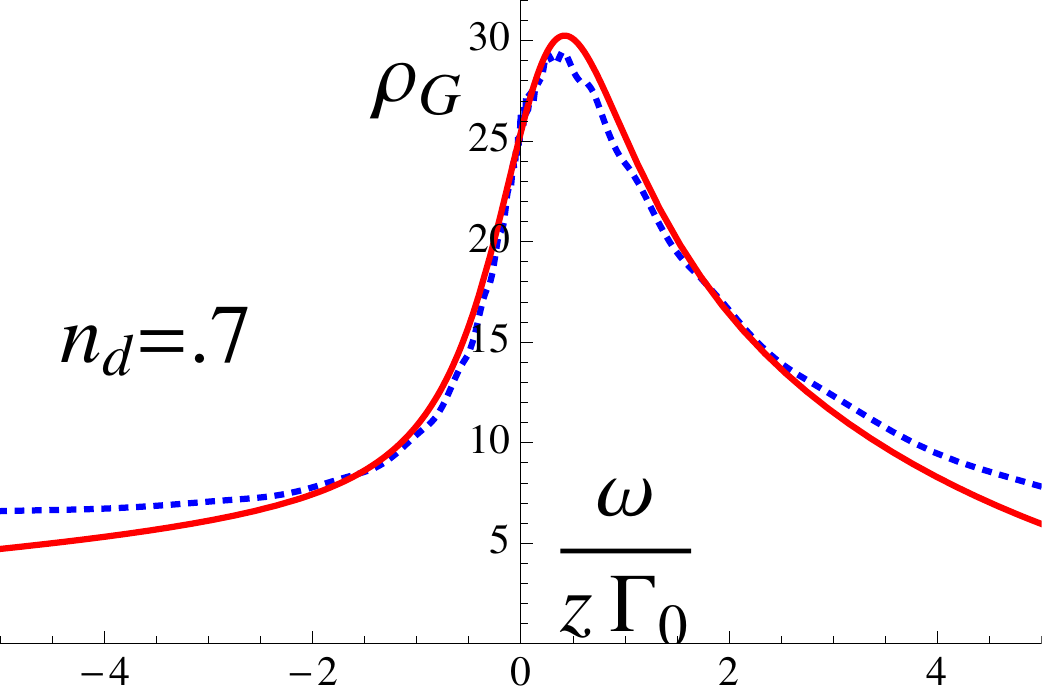}
\includegraphics[width=.3\columnwidth] {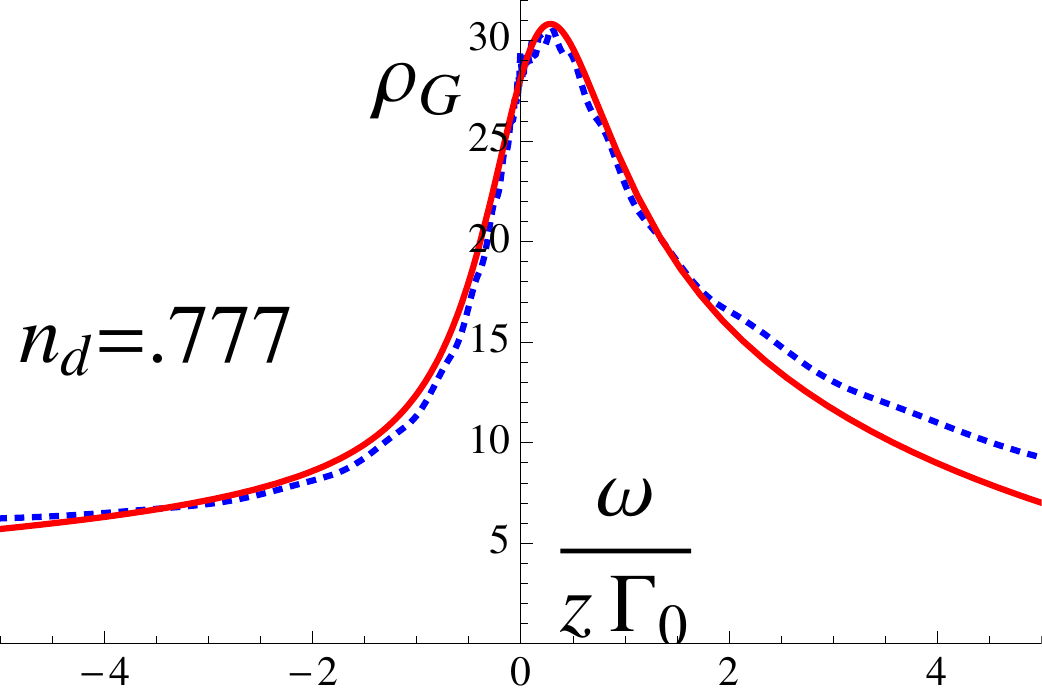}
\includegraphics[width=.3\columnwidth] {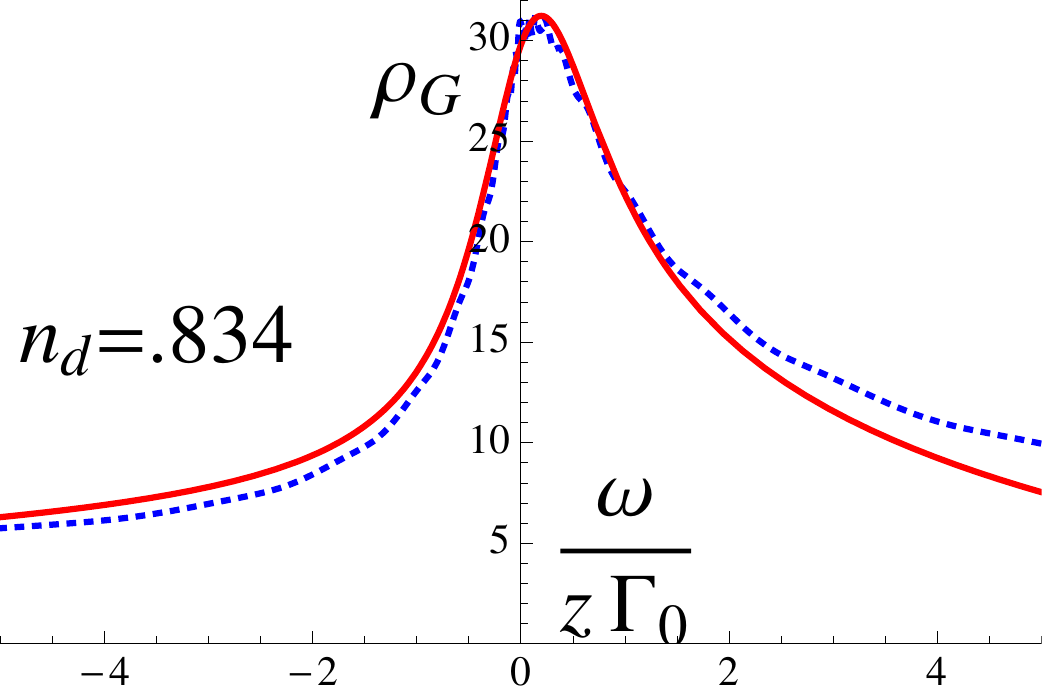}
\end{center}
\caption{The spectral density for the physical Green's function versus
  $\frac{\omega}{\Gamma_0 z}$ for densities of $n_d=0.35, 0.441, 0.6, 0.7,
  0.777,   0.834$. The red curve is the ECFL calculation, while the blue curve is the NRG calculation.}
\label{rhoGbefore}
\end{figure}
\end{widetext}

\begin{figure}[!h]
\begin{center}
\includegraphics[width=\columnwidth] {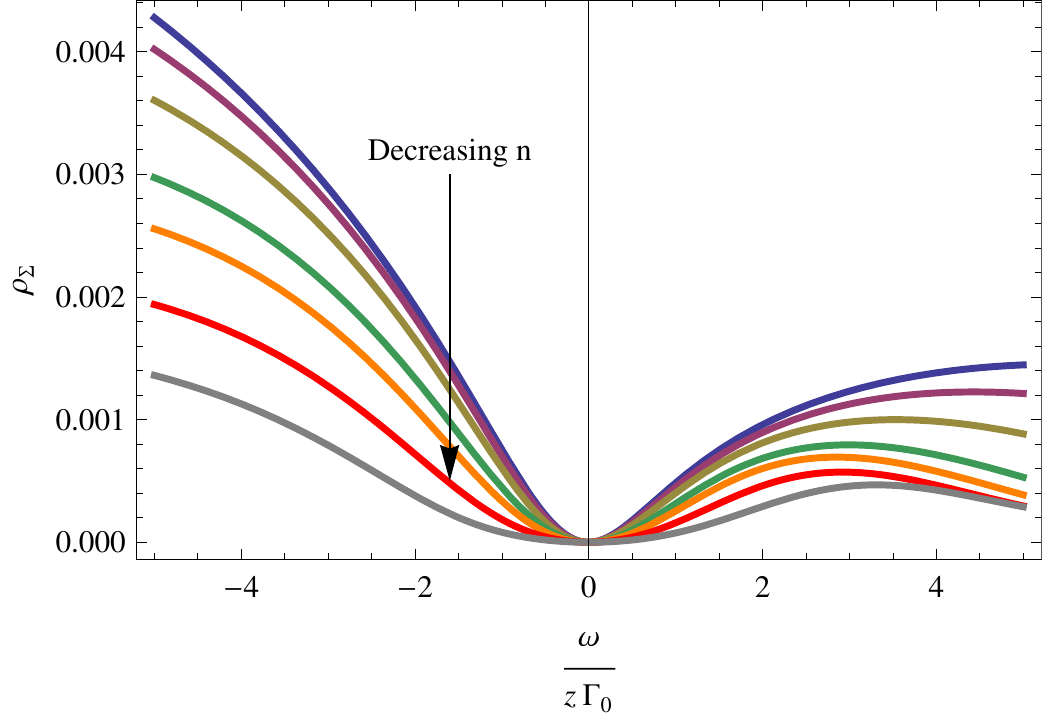}
\end{center}
\caption{The spectral function for the Dyson-Mori self-energy  for densities
  of $n_d=0.874, 0.777, 0.7, 0.6, 0.536, 0.441, 0.35.$ The curvature of the quadratic minimum becomes larger with increasing density. }
\label{rhoSigbeforel}
\end{figure}

  In \figdisp{rhoSigbeforel} we present the density evolution of the  spectral function for the Dyson Mori self-energy  (see \disp{DMSigma}). This exhibits a remarkable similarity to the analogous spectral density for the \tJ model in the limit of high dimensions \refdisp{agcoll} and the Hubbard model at large $U$ \refdisp{badmetal}. 

\section{Conclusion}

In this work we have applied the ECFL formalism at the simplest level, using the $O(\lambda^2)$ equations,  to the Anderson impurity model with
$U \rightarrow \infty$ . In this formalism, the two self-energies of the ECFL
theory $\Psi$ and $\chi$ are calculated using a skeleton expansion in the
auxiliary Green's function $\GH$. This is analogous to the skeleton expansion
 for the Dyson self-energy $\Sigma$, in  standard
Feynman-Dyson perturbation theory applicable to  the case of finite $U$. These two self-energies  determine $\GH$ as well as the physical $\G$, leading to a self-consistent solution. We obtained the
equations to second order and solved them numerically at $T=0$. We found that
at low enough $\omega$, the ECFL self-energies have symmetric spectra of the type predicted by
Fermi-Liquid theory (see \figdisp{rhoChibeforel}  and \figdisp{rhoPsibeforel}).
Combining them through the ECFL functional form \disp{DMSigma} generates a
non-trivial   self-energy with an asymmetric spectrum displayed in
\figdisp{rhoSigbeforel}.  It  therefore appears that functional form  \disp{DMSigma} has the potential  to generate realistic and non trivial spectral densities, starting with rather simple components. The availability of convenient and natural analytical expressions is seen to provide a distinct advantage of the ECFL formalism.  Formally exact techniques such as the NRG involve  steps  that are not not automatically endowed with these, but rather rely on analytic continuation  or other equivalent techniques.

The physical   spectral function for the impurity site is obtained  from the above pair of ECFL self energies, and displays a Kondo or  Abrikosov-Suhl resonance. This feature becomes more narrow and the spectrum becomes more skewed towards the  occupied side of the peak with increasing density. However, the computed quasiparticle  $z$ in the present calculation is larger than the exact value, we comment further on this aspect below.

 The location of the peak is set by $\epsilon_d +
\Sigma_{DM}(0)$ (\disp{DMG}). Using \disp{fsr-2}, we can see that this
quantity must decrease with increasing density. This is consistent with our
observation that the peak shifts to the left with increasing density. We
expect that the location of the peak will approach $\omega=0$ as $n_d
\rightarrow 1$. This can also be understood from the need to have more
spectral weight to the left of $\omega=0$ to yield a higher value of $n_d$.
We found that the ECFL spectrum satisfies the Friedel sum rule (\disp{fsr-1}) to
a high degree of accuracy, and  that ECFL yields values of $\epsilon_d$
 in good agreement with the NRG values at all densities (See
Table~(I).) 

As mentioned above the ECFL calculation to $O(\lambda^2)$  overestimates the value of the quasiparticle weight $z$ as
compared with the  NRG  and the exact asymptotic result $z\propto
e^{-1/2(1-n_d)}$ as $n_d\to 1$ \refdisp{rh84}, the difference becoming more significant with increasing density. This also leads to an over broadening of the peak in the ECFL spectrum at higher densities. This is consistent with the fact that the $\lambda$ expansion of the ECFL is a low-density expansion and the current calculation has only been carried out to $O(\lambda^2)$. Nevertheless, after rescaling the $\omega$ axis for both the ECFL and NRG spectra by their respective values of $z$, we find good quantitative agreement between the two as in \figdisp{rhoGbefore}.  In \figdisp{rhoGscale} we illustrate the comparison between scaled and unscaled spectral functions at a typical density. We  find similarly good agreement with the NRG calculation from \refdisp{Costi}.

Finally we note that the computed spectral functions exhibit a remarkable similarity to the analogous spectral density for the \tJ model in the limit of high dimensions \refdisp{agcoll} and the Hubbard model at large $U$ \refdisp{badmetal}.

\section{Acknowledgements} This work was supported by DOE under Grant No. FG02-06ER46319. We thank H. R. Krishnamurthy  for helpful comments.

\appendix
\section{Appendix A: Calculating the self-energies in the $O(\lambda^2)$ theory \lab{AppendixA} }
The calculation follows the procedure given in \refdisp{Monster}. A few comments are provided to make the connections explicit- the zeroeth order vertices are common to \refdisp{Monster}  Eqs.~(B3,\ B14), and  the first order $\U$ is common to Eq.~(B15). The first order vertex $[\Lambda]_1$ can be found parallel to Eq-(B23- B28)  from differentiating \beq[\GHI (i,f)]_1= \Delta(i,f) .\GH^{(k)}(i,i) + \delta(i,f) \Delta(i,\bb{a}). \GH^{(k)}(\bb{a},f),\eeq as
 \barray
 &&  [\Lambda^{(a)}(i,m;j)]_1  = - 2 \Delta(i,m). \GH(i,j). \GH(j,i) 
\nn \\
&&  -  2 \delta(i,m) \Delta(i,\bb{k}).\GH(\bb{k},j). \GH(j,i).\earray Here the bold labels are integrated over.   From this we construct the time domain self-energies 
\beq
\Psi(i,f)= - 2 \lambda \Delta(i,\bb{k}).\GH(\bb{k},f). \GH(i,f).\GH(f,i), \label{timedomain-1} \eeq 
and 
\barray
&&\Phi(i,f)= - \delta(i,f) \Delta(i \bb{k}).\GH(\bb{k} i) \nn \\
&& - 2 \lambda \Delta(i \bb{j}). \GH(\bb{j} \bb{k}). \Delta(\bb{k} f). \GH(\bb{k} i). \GH(i \bb{k}) \nn \\
&&- 2 \lambda \Delta(i \bb{j}). \GH(\bb{j} f). \Delta(f \bb{k} ). \GH(\bb{k} i). \GH(i f) . 
\label{timedomain-2}
\earray After shifting $\Delta(i,f)\to \Delta(i,f) - \frac{u_0}{2}\delta(i,f)$ and  Fourier transforming  these we obtain \disp{glambda} and \disp{chem}. These expressions for the self-energies are correct to $O(\lambda)$ and lead to expression for $\GHI$ and $\mu$ which are correct to $O(\lambda^2)$. $\chi$ can be extracted from $\Phi$ as indicated in the text.

\section{Appendix B: Frequency summations \lab{AppendixB}}
An efficient method to perform the frequency sums is to work with the time
 domain formulas \disp{timedomain-1} and \disp{timedomain-2} until the final
 step where Fourier transforms are taken. We note the representation for the
 Green's function
\beq
\GH(\tau) = \int_x \rho_\GH(x) e^{- \tau x} \left[ \theta(-\tau) f(x) - \theta(\tau) \bar{f}(x)\right],
\eeq
so that we can easily compound any pair that arises by dropping the cross products $\theta(\tau) \theta(-\tau)$ and using $\theta(\tau)^2= \theta(\tau)$.     An example illustrates this procedure:
\barray
\GH(\tau) \GH(-\tau) &=& -  \int_{x,y}  \rho_\GH(x) \rho_\GH(y) e^{- \tau (x-y)} \times
\nn \\
&& \left[ \theta(-\tau) f(x)\bar{f}(y) + \theta(\tau) \bar{f}(x) f(y)\right]. \label{eq-48}
\earray
We also need to deal with the convolution of pairs of functions.
\barray
X(\tau)& =& \int_{- \beta}^\beta \ d \bar{\tau} \ \GH(\bar{\tau}) \left[ \Delta(\tau-\bar{\tau} )- \frac{u_0}{2} \delta(\tau-\bar{\tau}) \right] \nn \\
&=& \int_x \rho_M(x) e^{- \tau x} \left[ \theta(-\tau) f(x) - \theta(\tau) \bar{f}(x)\right], \label{eq-49}
\earray
where the density $\rho_M(x)$ is defined in \disp{composite}. This equation in turn is easiest to prove by transforming into a product in the Matsubara frequency space,  simplifying using partial fractions, and then transforming back to time domain. 
We next note that \disp{timedomain-1} and  \disp{timedomain-2} imply
 \barray \Psi(\tau) &=& - 2 \lambda \  X(\tau) \GH(\tau) \GH(-\tau),  \nn \\
  \chi(\tau)& =& -2 \lambda \ X(\tau) X(-\tau) \GH(\tau),
  \earray
  so that taking Fourier transforms is simplest if first multiply out as in \disp{eq-48}, leading to \disp{selfenergies}.


\begin{thebibliography}{235}
\bibitem{fn-2} To recover \disp{fsr-1}, we may  use \disp{DMG} and  the Fermi liquid assumption of $\Im m \ \Sigma_{DM}(0)=0$ so that
$
\rho_{\G}(0) = \frac{1}{\pi} \frac{\Gamma_0 (1- \frac{n_d}{2})^2}{\Gamma^2_0 (1- \frac{n_d}{2})^2+ (\epsilon_d + \Re e \Sigma_{DM}(0))^2}, $
and combine with  \disp{fsr-2}.
\end{thebibliography}
\end{document}